\documentclass[twocolumn]{aastex631}
\graphicspath{{./}{Figures/}}
\submitjournal{ApJL}

\begin{document}

\shorttitle{Decoupling drives radial shear dynamos}
\shortauthors{Cao et al.}
\title{Core-envelope decoupling drives radial shear dynamos in cool stars}

\correspondingauthor{Lyra Cao}
\email{cao.lyra@gmail.com}

\author[0000-0002-8849-9816]{Lyra Cao}
\affiliation{Department of Astronomy, The Ohio State University, Columbus, OH 43210, USA}

\author[0000-0002-7549-7766]{Marc H. Pinsonneault}
\affiliation{Department of Astronomy, The Ohio State University, Columbus, OH 43210, USA}

\author[0000-0002-4284-8638]{Jennifer L. van Saders}
\affiliation{Institute for Astronomy, University of Hawai'i, 2680 Woodlawn Drive, Honolulu, HI, 96822, USA}

\begin{abstract}

Differential rotation is thought to be responsible for the dynamo process in stars like our Sun, driving magnetic activity and star spots. We report that star spot measurements in the Praesepe open cluster are strongly enhanced only for stars which depart from standard models of rotational evolution. A decoupling of the spin down history between the core and envelope explains both the activity and rotation anomalies: surface rotational evolution is stalled by interior angular momentum redistribution, and the resultant radial shears enhance star spot activity. These anomalies provide evidence for an evolving front of shear-enhanced activity affecting the magnetic and rotational evolution of cool stars and the high-energy environments of their planetary companions for hundreds of millions to billions of years on the main sequence.

\end{abstract}


\section{Introduction} \label{sec:intro}
Magnetic fields are ubiquitous in cool stars. Young stars have significant X-ray fluxes \citep{1994ApJS...91..625S} and emission features \citep{1963ApJ...138..832W} in the cores of strong lines that provide evidence for magnetically driven non-thermal chromospheric and coronal heating of their outer atmospheres. Stellar activity diagnostics are variable on moderate to long timescales for stars like our Sun \citep{1998ApJS..118..239R}, which supports a dynamo origin for the magnetic fields that generate them \citep{1993ApJ...408..707P}. The standard interface dynamo model for the Sun relies on a radiatively stable core where differential rotation can amplify fields at the base of the turbulent convection zone \citep{1993ApJ...408..707P, 1997ApJ...486..502C}. Alternative dynamo modes have been postulated but not observed \citep{1997ApJ...486..502C}, as stars less massive than ~0.35 solar masses (lacking an interface) appear to follow the same activity scaling relations as more massive stars \citep{2011ApJ...743...48W}. Dynamos driven by surface convection launch magnetized winds, which are efficient engines for extracting angular momentum \citep{1967ApJ...148..217W}, thus linking observed stellar rotation rates to dynamo action. Perturbations in our standard dynamo framework therefore also impact gyrochronology, the use of stellar rotation as an age diagnostic \citep{2007ApJ...669.1167B}.

Standard magnetic braking models have been unsuccessful in explaining recent observations of low-mass stars. Surface rotation evolution is observed to stall, but then resume, in a strongly mass-dependent fashion \citep{2020ApJ...904..140C}. The epoch of stalling occurs at tens of Myr for solar analogs and hundreds of Myr for lower mass stars \citep{2020A&A...636A..76S}. Apparent rotational stalls with ongoing angular momentum loss can be explained with a lag time between the spin down of the outer turbulent layers and the inner radiative core, which is referred to as core-envelope decoupling \citep{1991ApJ...376..204M}. Decoupling implies the temporary presence of strong radial shears below the surface convection zone, which could induce a distinct departure from standard braking laws.

In this paper we demonstrate that stars experiencing stalled surface spin down also have unusually strong surface magnetic fields. In Praesepe, these stars have spot levels comparable to saturation, which is a maximal state of surface magnetism. This provides evidence for an alternative dynamo mechanism driven by radial shears, supports core-envelope decoupling as an astrophysical phenomenon, and demonstrates that stars in the epoch of decoupling experience shear-enhanced magnetism for hundreds of Myr to Gyr's on the main sequence.

\section{Motivation \& Methods} \label{sec:methods}

Prior attempts to explain rotational stalling with core-envelope decoupling have not been definitive. Core-envelope decoupling gyrochrones appear to reproduce the morphology of open cluster rotation trends with mass and age \citep{2020A&A...636A..76S} but cannot rule out a temporary weakening of stellar winds or surface magnetism as the source of the stalling \citep{2020ApJ...904..140C}. Rotationally induced mixing can also explain main sequence lithium depletion. Evolutionary models of lithium depletion favor larger internal shears for longer timescales in lower mass stars, consistent with core-envelope decoupling predictions \citep{2016ApJ...829...32S}; however, such models assume an invariant birth rotation distribution between clusters, and deviate from observations at still older ages. Nevertheless, recent ground-based observations of field stars observe that a gap in the period distribution of more massive stars disappears in fully convective stars \citep{2022AJ....164..251L}, which is expected if core-envelope decoupling is the source of the feature.

If core-envelope decoupling actually occurs, we expect strong internal shears during that epoch to have a direct observational signature. The large rotational shear between stellar surface and core predicted by decoupling is thought to drive dynamo action \citep{1997A&A...326.1023B}. This contrasts with the standard interface dynamo model, which predicts that stars with slow surface rotation would be inactive. However, a conclusive enhanced activity signal has not previously been seen in the literature, we argue largely because of poor sampling and precision. Starspot measurements have been documented as overactive in a handful of young stars in the Pleiades but are limited to an unconverged rotational sequence \citep{2022MNRAS.517.2165C}. UV measurements suggested overactivity in a small sample of K dwarfs in the Hyades, but with considerable scatter \citep{2022ApJ...929..169R}.

A technique for measuring starspot filling factors from APOGEE high-resolution near-infrared spectra was recently published \citep{2022MNRAS.517.2165C}, introducing a population-scale technique for studying stellar magnetism. The ensemble of spectral features in the H-band permit precision measurements of starspots with typical dispersions ranging from $\sim$6.6\% in active stars to $\sim$1.1\% in solar type stars, which enabled the construction of a mean relation between Rossby number (the ratio of rotation period to convective overturn timescale) and starspots \citep{2022MNRAS.517.2165C}. The recent 17th Data Release of the APOGEE survey \citep{2021AJ....162..302B} includes a deep sampling of the $\sim$670 Myr \citep{2019ApJ...879..100D} anomalously rotating open cluster Praesepe, enabling a detailed study of stellar magnetism during the epoch of stalled spin down.

\subsection{Sample Selection and Stellar Parameters}\label{sec:sample}
The open cluster samples in this work are single Praesepe members with APOGEE DR17 spectra and measured rotation periods. We use members of an open cluster so that our sample stars have well-known masses, compositions, and ages. The Praesepe open cluster is well-studied in the literature for both membership and binarity. We aggregate 2542 members out of 2902 systems from Gaia DR2 \citep{2018A&A...616A..10G, 2018A&A...618A..93C, 2019MNRAS.486.5405G, 2019A&A...627A...4R, 2019A&A...628A..66L, 2021ApJ...923..129J, 2021ApJS..257...46G} and literature \citep{2007AJ....134.2340K} sources. The APOGEE spectra for Praesepe were obtained as part of the Bright Time Extension program \citep{2021AJ....162..302B}, which targeted Gaia DR2 literature members, prioritizing low mass main sequence stars down to $B_P-R_P \sim 3.3$. Selected APOGEE targets were observed to a target signal-to-noise ratio above 100 per pixel; this resulted in 375 member spectra in Praesepe.

Two-temperature spectroscopic fits were performed with the least-squares solver {\sc{ferre}} on high-resolution (R$\sim$22,500) near-infrared H-band spectra \citep{2022MNRAS.517.2165C}. This jointly provides measurements of starspot filling fraction ($f_{\mathrm{spot}}$), starspot temperature contrast ($x_{\mathrm{spot}}$), effective temperature ($T_{\mathrm{eff}}$), surface gravity (logg), rotational velocity ($v\,\sin\,i$), metallicity ([M/H]), and microturbulence ($v_{\mathrm{dop}}$). Our method is applicable to stars with $T_{\mathrm{eff}}$ between 3400--6200 K. We therefore restricted our sample to targets in the color range $0.7 < B_P-R_P < 2.5$, leaving 204 stars. As a quality cut, we removed stars with spectroscopic signal-to-noise ratios below 70 per pixel, corresponding to the ASPCAP {\texttt{SN\_WARN}} flag \citep{2021AJ....162..302B}, which we found correlated with significantly increased uncertainty in $f_{\mathrm{spot}}$. This left 179 stars with robust spectroscopic solutions, a recovery fraction of 88\% (179/204).

We also required rotation periods for gyrochronological analysis. Our source of rotation periods \citep{2017ApJ...839...92R} used space-based K2 data and had nearly complete recovery for open cluster stars, consistent with having missed only targets seen nearly pole-on without a rotation modulation signal. Our final target sample, before binary rejection, contained 136 stars in Praesepe---for a total completeness of 67\% (136/204) limited mainly by APOGEE targeting \citep{2021AJ....162..302B}. The resulting sample in Table \ref{tab:clusterdata} is representative of literature cluster members, with no gaps in the cluster sequence resulting from this selection. Binaries, which are contaminants in spectroscopic starspot measurements, are comprehensively rejected as described in Appendix \ref{sec:binaryrej}.

\begin{table}
\centering
\caption{Column descriptions for the open cluster starspot table. Spectroscopic measurements from two-temperature fits are published with rotation data for the reported Praesepe sample and the prior Pleiades calibration sample.} \label{tab:clusterdata}
\begin{tabular*}{\columnwidth}{@{}l@{\hspace*{5pt}}l@{}}
  \hline
  Label & Contents \\
  \hline
  APOGEE\_ID & Source 2MASS ID \\
  Cluster & Praesepe or Pleiades \\
  Teff & Two-temperature effective temperature (K) \\
  e\_Teff & Uncertainty in derived Teff \\
  Mass & Inferred mass from spotted models \\
  fspot & Two-temperature starspot filling fraction \\
  e\_fspot & Uncertainty in derived fspot \\
  xspot & Two-temperature starspot temp. contrast \\
  e\_xspot & Uncertainty in derived xspot \\
  logg & Two-temperature surface gravity \\
  e\_logg & Uncertainty in derived logg \\
  {[}M/H{]} & Two-temperature metallicity \\
  e\_{[}M/H{]} & Uncertainty in derived [M/H] \\
  vsini & Two-temperature rotational velocity (km/s) \\
  e\_vsini & Uncertainty in derived vsini \\
  vdop & Two-temperature microturbulence (km/s) \\
  e\_vdop & Uncertainty in derived vdop \\
  Period & Rot. period from \citet{2016AJ....152..113R, 2017ApJ...839...92R} \\
  Notes & B: Flagged as binary, O: Over-spotted \\
\hline
\multicolumn{2}{l}{This table is entirely available in machine-readable form.}
\end{tabular*}
\end{table}

\subsection{Gyrochronology Models} \label{sec:gyrochrones}
To model rotational evolution in stars we take stellar evolution calculations and (1) define their initial angular momenta, (2) characterize the angular momentum losses from their outer envelopes due to magnetized winds, and (3) model their internal angular momentum transport. Rotational models are launched from the $\sim$10 Myr Upper Sco rotational sequence \citep{2018AJ....155..196R}, by which age massive accretion disks which exchange angular momentum with the host star are nearly absent. We initialize stellar models which account for the structural impact of starspots \citep{2020ApJ...891...29S} between the 7.5--92.5th percentiles in rotation for Upper Sco to capture the range in observed rotation with minimal contamination (Appendix \ref{sec:gyro}).

We adopt an angular momentum loss prescription for magnetized winds \citep[Equation 1 from][]{2013ApJ...776...67V} with two free parameters relating to the torque: $\omega_{\mathrm{crit}}$, the saturation threshold above which the torque transitions from scaling at dJ/dt $\sim$ $\omega^3$ to dJ/dt $\sim$ $\omega$, and $f_k$, the overall normalization of the braking torque.
To model internal angular momentum transport, we use a standard two-zone model where angular momentum is exchanged from a radiative core to a convective envelope over a characteristic coupling timescale \citep[Equations 13--15 from][]{2010ApJ...716.1269D}. We track changes in the extent of the convection zone with evolutionary models, and treat the core-envelope coupling timescale as a free parameter. We fit a standard mass-dependent scaling relation to open clusters of the form $\tau_{\mathrm{ce, \,} \star} = \tau_{\mathrm{ce, \,} \odot} \left( M / M_\odot \right)^{\alpha}$ \citep{2015A&A...584A..30L,2016ApJ...829...32S,2020A&A...636A..76S}, where $\tau_{\mathrm{ce, \,} \odot}$ is the solar coupling timescale and $\alpha$ is the mass-dependent power law exponent.

We jointly calibrate these four parameters of the wind law ($\omega_{\mathrm{crit}}$, $f_k$) and two-zone model ($\tau_{\mathrm{ce, \,} \odot}$, $\alpha$) to reproduce the Pleiades and Praesepe rotational distributions, as well as the rotation of the Sun, from an Upper Sco launch. This calibration captures the mean rotational evolution and the variation between rapid and slow rotators. Our approach, described in detail in Appendix \ref{sec:gyro}, differs from standard practice in that we independently incorporate the structural impact of both magnetism and metallicity, simultaneously fitting the Pleiades (112 Myr, $\mathrm{{[M/H]}}=0$) and Praesepe (670 Myr, $\mathrm{{[M/H]}}=+0.188$).

\section{Results \& Analysis} \label{sec:results}

\begin{figure*}
\centering
\includegraphics[trim={0cm 0cm 0cm 0cm},width=0.8\linewidth]{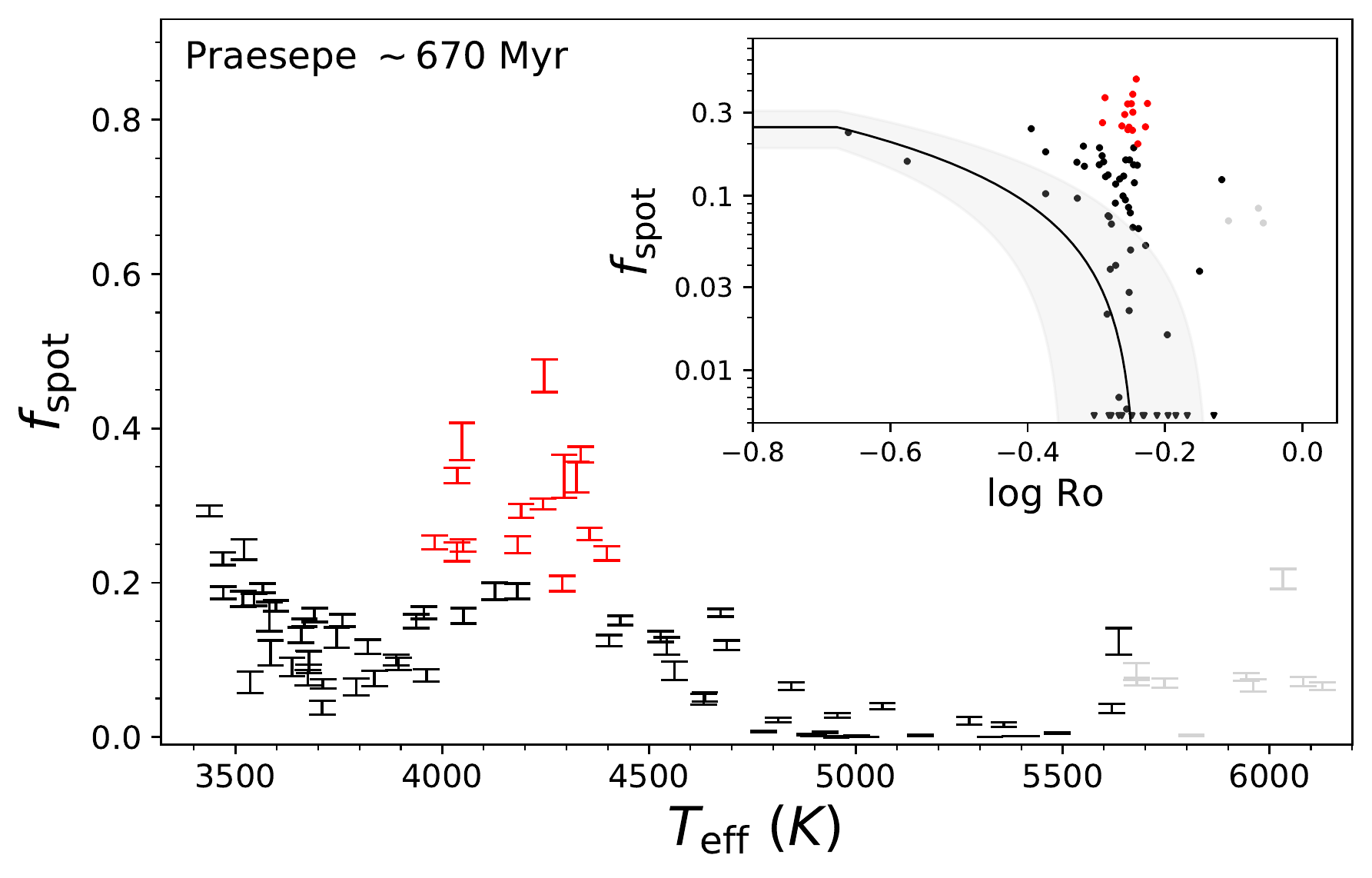}
\caption{\label{fig:RadialShears_fspotTeff} Starspot filling fractions derived from the fit to APOGEE DR17 spectra are reported for single stars with rotation periods in Praesepe. Red errorbars are $3\sigma$ outliers in the Ro--fspot relation \citep[Equation 4 from][]{2022MNRAS.517.2165C} after accounting for total population dispersion, including inclination and cyclic variation effects. Gray errorbars are stars in the same color range as possibly spuriously overactive stars in M67 \citep{2022MNRAS.517.2165C} and should be interpreted with caution. Over-active stars in Praesepe are tightly clustered; 15 of 21 stars between $3950 < T_{\mathrm{eff}} < 4410$ K are $> 3 \sigma$ discrepant in activity. Inset: the $f_{\mathrm{spot}}$--Rossby diagram is shown with upper limits represented as triangles.}
\end{figure*}

\begin{figure*}
\centering
\includegraphics[trim={0cm 0cm 0cm 0cm},width=0.8\linewidth]{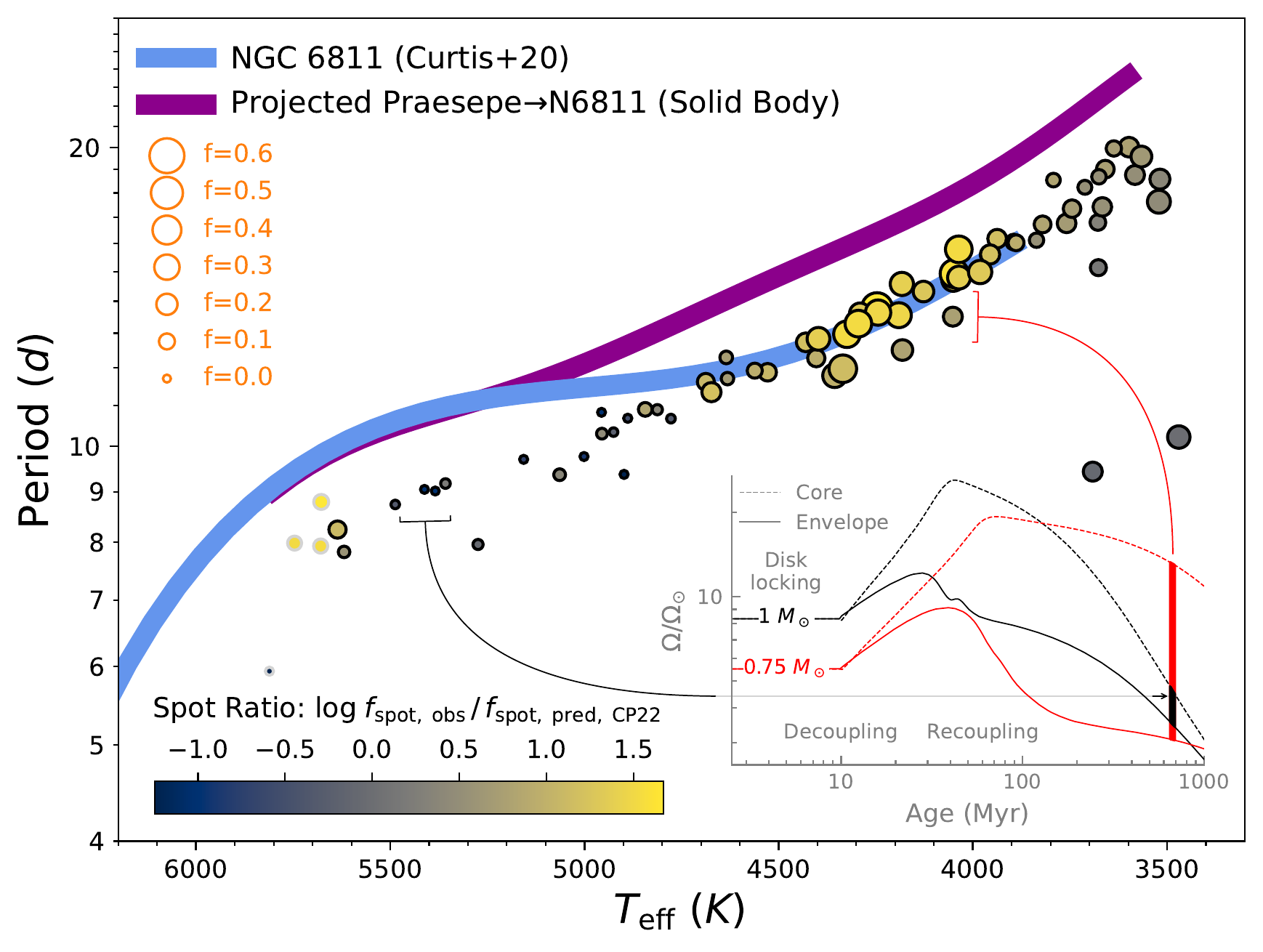}
\caption{\label{fig:RadialShears_PeriodTeff} The Praesepe rotation period distribution, indicating observed rotation and spot anomalies. Symbol sizes are scaled to the measured starspot filling fraction and colors correspond to the logarithm of the ratio between the observed spot fraction and predictions from a standard Rossby scaling \citep[Equation 4 from][]{2022MNRAS.517.2165C}. The blue line indicates a fit to NGC 6811 (1 Gyr) after transforming temperatures from \citet{2020ApJ...904..140C} to our spotted temperature system using Praesepe stars in common. Points lying on the blue line appear to stall in their surface spin down between the ages of Praesepe and NGC 6811. The purple line instead indicates the expected rotation periods of these stars by NGC 6811 age assuming solid-body rotation. Inset: the rotational evolution of cores (dashed) and envelopes (solid) for spotted models at 1.0 and 0.8 $M_\odot$, showing large radial shears in the 0.8 $M_\odot$ model at Praesepe age.}
\end{figure*}

We measure starspot filling fractions and temperature contrasts jointly with spectroscopic stellar parameters ($T_{\mathrm{eff}}$, [M/H], logg, $v\,\sin\,i$, $v_{\mathrm{dop}}$) in 179 Praesepe members, obtaining a final sample of 88 single stars with rotation periods (Appendix \ref{sec:binaryrej}). Figure \ref{fig:RadialShears_fspotTeff} shows the dependence of starspot filling factor with stellar effective temperature. Stars highlighted in red are overspotted for their Rossby number by $> 3\sigma$ after accounting for total cycle and inclination variation, found empirically to be 6.0\% in Praesepe. Assuming starspot filling fractions are normally distributed about the mean Rossby relation, the probability of finding $\geq$ 15 stars to be 3$\sigma$ discrepant out of 88 single stars is 0.00003\%. Combined long-term and short-term starspot cycle variations estimated using scaling relations inferred from the Mt. Wilson solar analog \citep{1998ApJS..118..239R} and starspot cluster samples \citep{2022MNRAS.517.2165C} yield $f_{\mathrm{spot, \, rms}}$ $\sim$1.2\% for solar-type stars and $f_{\mathrm{spot, \, rms}}$ $\sim$0.9\% for active stars. As over-active stars here are discrepant by at least 18--20\% in $f_{\mathrm{spot}}$, the total cycle variation is an order of magnitude too low to explain these departures. Over-active stars are confined to a narrow mass range, making starspot cycles an unlikely explanation. These stars are confirmed members and pass checks against binary contamination.

In Figure \ref{fig:RadialShears_PeriodTeff} we show the 670 Myr Praesepe sequence as points scaled by $f_{\mathrm{spot}}$ and colored by the ratio between observed and predicted $f_{\mathrm{spot}}$ using an empirical starspot relation \citep{2022MNRAS.517.2165C}. We overlay these points on the 1 Gyr old cluster NGC 6811 in blue, showing that stars which have not spun down during the $\sim$300 Myr of evolution between Praesepe and NGC 6811 are significantly spot-enhanced.

Stellar activity is related to the predicted spin down rate in wind models, so the observed starspot enhancements disfavor models where stalled spin down is associated with a reduction in starspot activity.
Our data represent a striking counterexample to standard models of stellar dynamo and angular momentum evolution. No other proposed mechanism naturally produces the simultaneous magnetic and rotational anomalies in Praesepe without core-envelope decoupling. These stars are on the fast branch of the McQuillan gap \citep{2014ApJS..211...24M}, and their effective temperatures are affected by starspots \citep{2022MNRAS.517.2165C}, which suggests that the morphology of the gap may be influenced by a transition in surface spot coverage.

\section{Discussion \& Conclusion} \label{sec:disc}

\begin{figure*}
\centering
\includegraphics[trim={0cm 0cm 0cm 0cm},width=0.8\linewidth]{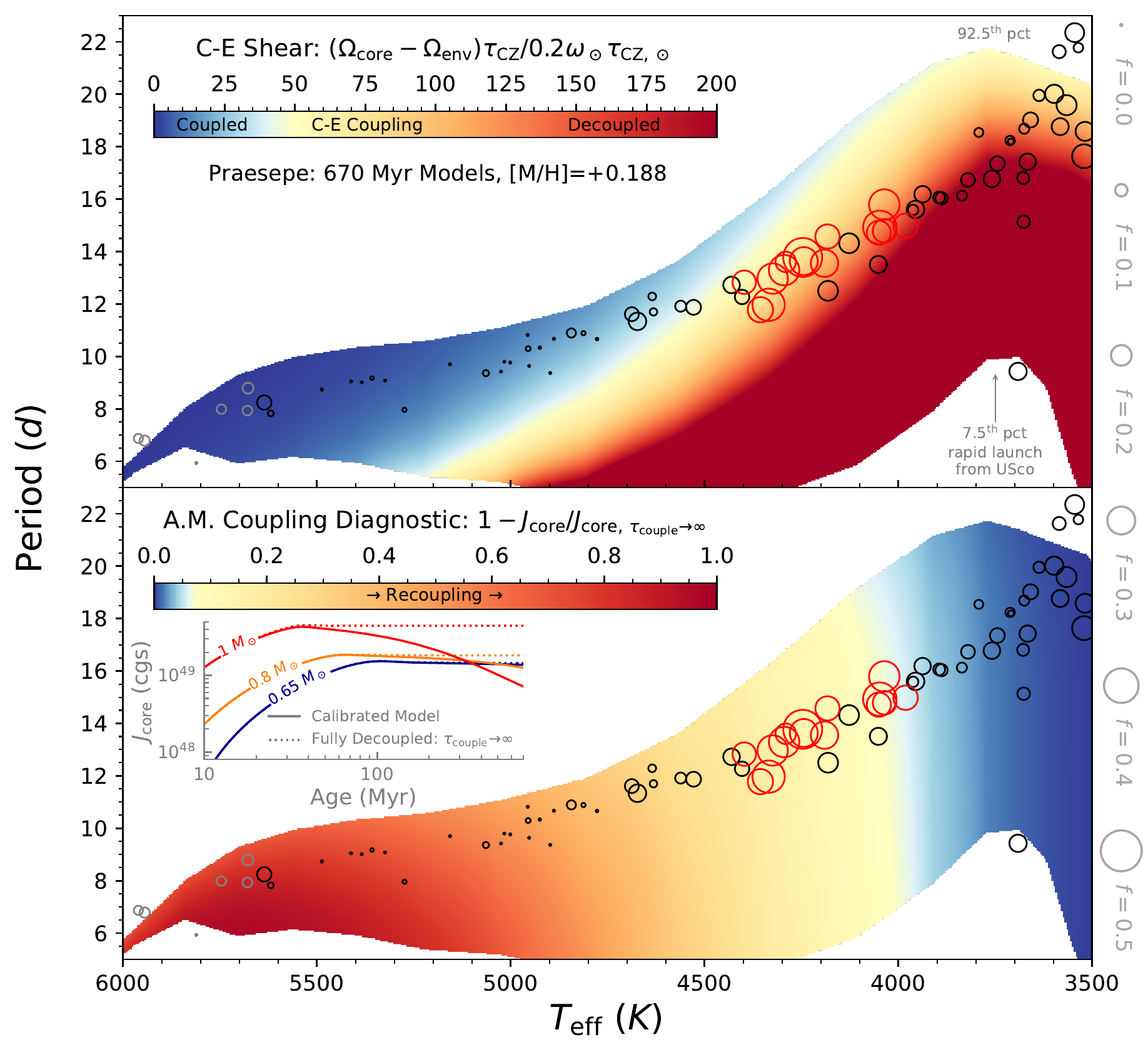}
\caption{\label{fig:RadialShears_ShearsAM} Top: core-envelope shear $S_{\mathrm{ce}}$, defined as the ratio between the radial shear and a solar latitudinal shear (assumed to be 0.2), represented in background colors. Bottom: the fractional change in the angular momentum content between cores of calibrated models and fully decoupled models $F_{\mathrm{couple}}$, a diagnostic for angular momentum transport. Circles are scaled in size to observed starspot fraction with the color convention from Figure \ref{fig:RadialShears_PeriodTeff}. Shear ratios greater than 200 on the top plot are represented as the same color to improve visibility. Overactivity is seen for models with both strong internal shears (top, $S_{\mathrm{ce}} \gtrsim 50$) and large integrated angular momentum fluxes (bottom, $F_{\mathrm{couple}} \gtrsim 0.05$). Inset: evolution of core angular momentum for 1.0, 0.8, and 0.65 $M_\odot$ fiducial rotational models (solid) and fully decoupled models (dotted). By Praesepe age, the two higher mass models depart from full decoupling and show significant angular momentum exchange, but the 0.65 $M_\odot$ model does not.}
\end{figure*}

\begin{figure*}
\centering
\includegraphics[trim={0cm 0cm 0cm 0cm},width=0.9\linewidth]{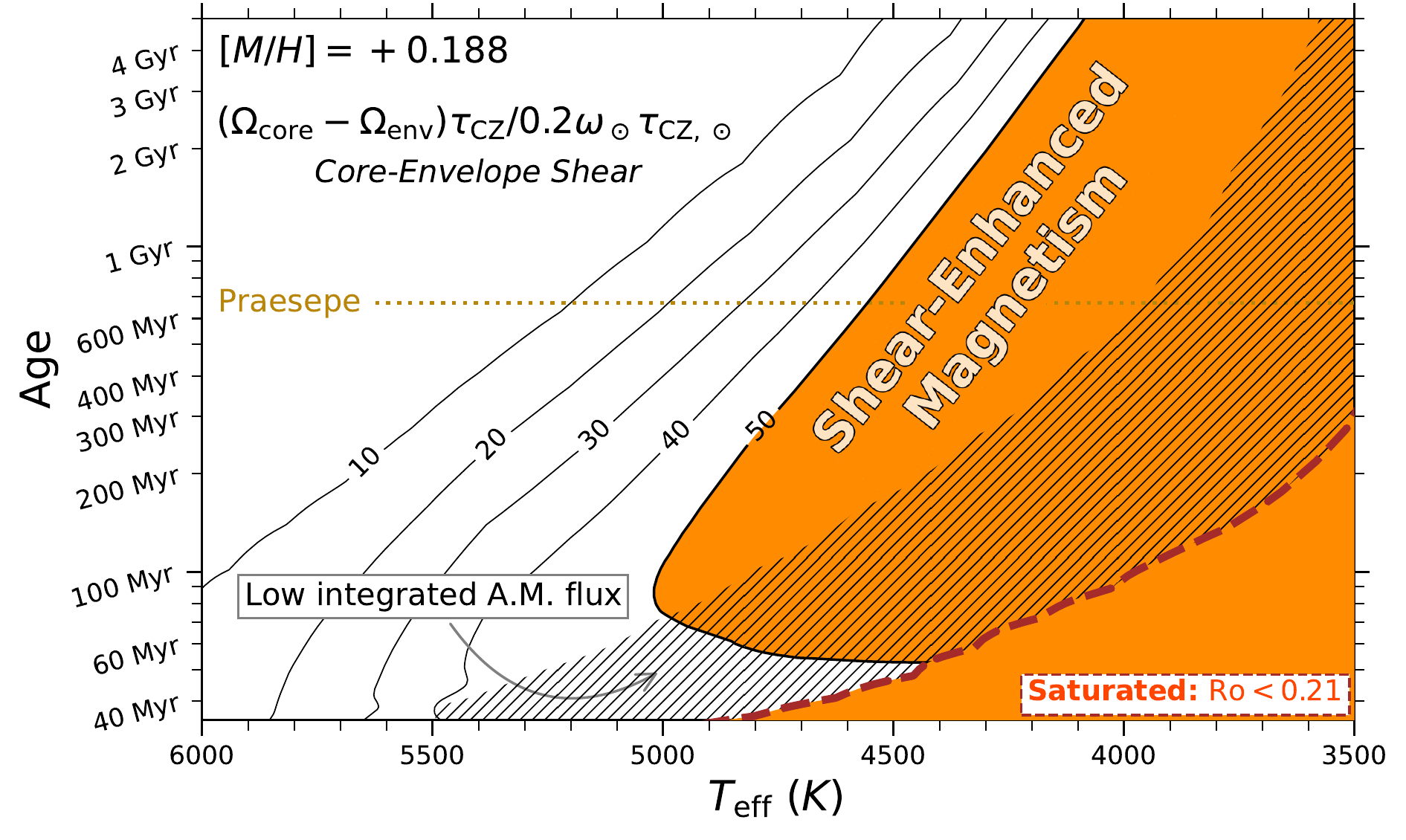}
\caption{\label{fig:RadialShears_ShearAge} The time evolution of core-envelope shear is plotted for models launched from the Upper Sco rotational sequence that reproduce the observed rotation at Praesepe age and metallicity. Contours of core-envelope shear ratios are shown ranging from 10--50. The hatched region corresponds to models where the integrated angular momentum flux from core to envelope is relatively weak (the A.M. coupling diagnostic is $\leq 0.05$). The region bounded by a brown dashed contour represents a standard starspot saturation criterion \citep{2022MNRAS.517.2165C}. The region with shear ratios $>50$ and significant angular momentum exchange is interpreted to be the locus of shear-enhanced magnetism in stars. The timescale for shear-enhanced magnetism varies strongly with $T_{\mathrm{eff}}$, from $\sim$100 Myr at 4900 K to $\sim$700 Myr by 4500 K, and $\sim$5 Gyr by 4100 K.}
\end{figure*}

Models initialized with rotation rates spanning the 7.5--92.5th percentiles in Upper Sco (see Appendix \ref{sec:gyro}) are plotted in Figure \ref{fig:RadialShears_ShearsAM}. In the top panel, we plot $S_{\mathrm{ce}}$, the core-envelope radial shear $(\Omega_{\mathrm{core}} - \Omega_{\mathrm{env}}) \tau_{\mathrm{CZ}}$, where $\tau_{\mathrm{CZ}}$ is the convective overturn timescale, normalized to the solar latitudinal shear. The observed overactivity in Praesepe coincides with transitions to high internal shears in two-zone rotational models, with the strongest signal seen at intermediate shears.

In the bottom panel of Figure \ref{fig:RadialShears_ShearsAM} we demonstrate the impact of recoupling on core angular momentum content. We introduce a coupling diagnostic, $F_{\mathrm{couple}} = 1-J_{\mathrm{core}}/J_{\mathrm{core, \, } \tau_{\mathrm{couple}} \to \infty}$, where $J_{\mathrm{core}}$ denotes the angular momentum of the stellar core and $J_{\mathrm{core, \, } \tau_{\mathrm{couple}} \to \infty}$ to the same in a corresponding decoupled model with recoupling switched off. We observe that stars with strong shears ($S_{\mathrm{ce}} \gtrsim 200$) but limited A.M. transport ($F_{\mathrm{couple}} \lesssim 0.05$) from the core are not overactive, while stars with intermediate shears ($200 \gtrsim S_{\mathrm{ce}} \gtrsim 50$) and ongoing A.M. transport ($F_{\mathrm{couple}} \gtrsim 0.05$) are over-spotted.

Identifying significant internal shears and core-envelope coupling as parameters correlated with shear-enhanced activity in Praesepe, we model the time-evolution of shear-enhanced magnetism at fixed rotation and metallicity in Figure \ref{fig:RadialShears_ShearAge}. Core-envelope shears enhance activity in K dwarfs in Praesepe, but at older ages the magnetic anomaly moves to lower mass stars (Figure \ref{fig:RadialShears_ShearAge})---a consequence of a coupling timescale varying strongly with stellar mass \citep{1991ApJ...376..204M,2016ApJ...829...32S,2010ApJ...716.1269D,2015A&A...584A..30L,2015A&A...577A..98G}. Only unusually rich young clusters deviate from standard rotational distributions \citep{2013A&A...560A..13M} and the structural impact of metallicity appears to dominate observed trends with rotation and activity \citep{2020MNRAS.499.3481A, 2021ApJ...912..127S}, suggesting that this result is robust. We predict starspot enhancements in Ruprecht 147 \citep[$\sim$ 2.7 Gyr, ][]{2020ApJ...904..140C} between $\sim$3800--4300 K for a [M/H] of +0.1, and in M67 ($\sim$4 Gyr) between $\sim$3700--4300 K at solar metallicity, noting that a metallicity offset of +0.1 dex roughly decreases the $T_{\mathrm{eff}}$ at fixed mass by $\sim$100 K. This work is evidence for an evolving shear front affecting cool stars, differentially enhancing stellar surface magnetism and impacting their high-energy radiation environments for hundreds of Myr to Gyr during the recoupling phase.

The large starspot filling fractions measured in Praesepe over-active stars are not merely a general feature of K dwarf spectra; in the Pleiades, there is no activity enhancement in K dwarfs \citep{2022MNRAS.517.2165C}. In the $\sim$4 Gyr old open cluster M67, early K dwarfs appear unspotted, with filling fractions consistent with zero \citep{2022MNRAS.517.2165C}. Though sparsely sampled by DR17, there also appears to be a spot enhancement signal for K dwarfs in the $\sim$727 Myr \citep{2019ApJ...879..100D} Hyades cluster. Over-active stars have similar spectroscopic chi-squared values to other members, and do not preferentially show convergence failures in spectroscopic fits. The contamination rate of unresolved binaries is minimal due to our comprehensive binary rejection (Appendix \ref{sec:binaryrej}). APOGEE targets members to lower masses than our sample \citep{2021AJ....162..302B}, the rotation catalog \citep{2017ApJ...839...92R} is highly complete ($\sim$67\%), and starspot measurements are precise for stars both more and less active than our sample \citep{2022MNRAS.517.2165C}. In the range we identify for overactivity in Praesepe ($1.5 < B_P-R_P < 1.9$, or $3950<T_{\mathrm{eff}} < 4410$ K), all 21 single stars, and all 9 members flagged as binaries, have spot fractions $> 1\sigma$ above the mean relation.

The mechanism mediating angular momentum transport between the core and envelope is actively debated. Hydrodynamic mechanisms cannot explain the internal solar rotation profile alone, so waves or magnetic fields are probable causes \citep{1997ARA&A..35..557P}.
Our observations disfavor models involving strong magnetic coupling as they predict rigid internal rotation on short timescales \citep{2019MNRAS.485.3661F}. Instead of direct coupling, models which exchange angular momentum non-magnetically between a magnetically-coupled core and convective envelope across thin shear layers permit long coupling timescales and persistent core-envelope shears \citep{2013ApJ...778..166O}. Simple estimates of the characteristic angular momentum transport timescale via internal gravity waves within 0.5 pressure scale heights of the convective zone boundary \citep[following section 4.2 of][]{2014ApJ...796...17F} produce sufficiently slow timescales which scale strongly ($\propto M^{-7.6}$) with stellar mass.
Alternatives to the decoupling model exist, such as the increase in complexity of magnetic field morphologies \citep{2018ApJ...862...90G}. Although interesting in the saturated domain, such mechanisms do not predict significant departures from the mean Rossby scaling and cannot explain the over-activity of stars experiencing stalled spin down. Magnetic field measurements in field populations do not appear to show a decline in the dipole field strength \citep{2019ApJ...886..120S}, but future work will more conclusively discriminate between these models.

We show in this work that core-envelope decoupling and stalled spin down are linked, producing observables not only in angular momentum evolution but also in activity. The existence of shear-enhanced magnetism suggests that stellar mixing processes can persist for hundreds of millions to billions of years---consistent with models where waves are the primary mechanism linking the cores and envelopes of stars. Evolutionary models will need to include core-envelope decoupling and angular momentum losses from dynamos with large core-envelope shears to obtain accurate rotation-based ages on the main sequence. Habitability models will need to be informed by the enhanced magnetism associated with core-envelope decoupling, as shear-enhanced activity can last for billions of years in low mass stars. Decoupling signals in activity and rotation are strongly mass- and age-dependent, and thus a potential age diagnostic for low-mass main sequence stars where other age indicators are scarce.

\begin{acknowledgments}
We thank the SDSS-IV collaboration for making this research possible. Funding for the Sloan Digital Sky Survey IV has been provided by the Alfred P. Sloan Foundation, the U.S. Department of Energy Office of Science, and the Participating Institutions. SDSS-IV acknowledges support and resources from the Center for High Performance Computing at the University of Utah. The SDSS website is www.sdss4.org.

L.C and M.H.P acknowledge computing resources and support made available by the Ohio State University College of Arts and Sciences.

L.C. and M.H.P. acknowledge support from TESS Cycle 5 GI program G05113 and NASA grant 80NSSC19K0597. J.v.S. acknowledges support from the National Science Foundation grant AST-1908723. This research was supported in part by the National Science Foundation under Grant No. NSF PHY-1748958.
\end{acknowledgments}

\bibliography{main}{}
\bibliographystyle{aasjournal}

\appendix
\section{Additional Methodology}
\subsection{Binary Rejection}\label{sec:binaryrej}

\begin{figure}[!thb]
\centering
\includegraphics[trim={0cm 0cm 0cm 0cm},width=0.8\columnwidth]{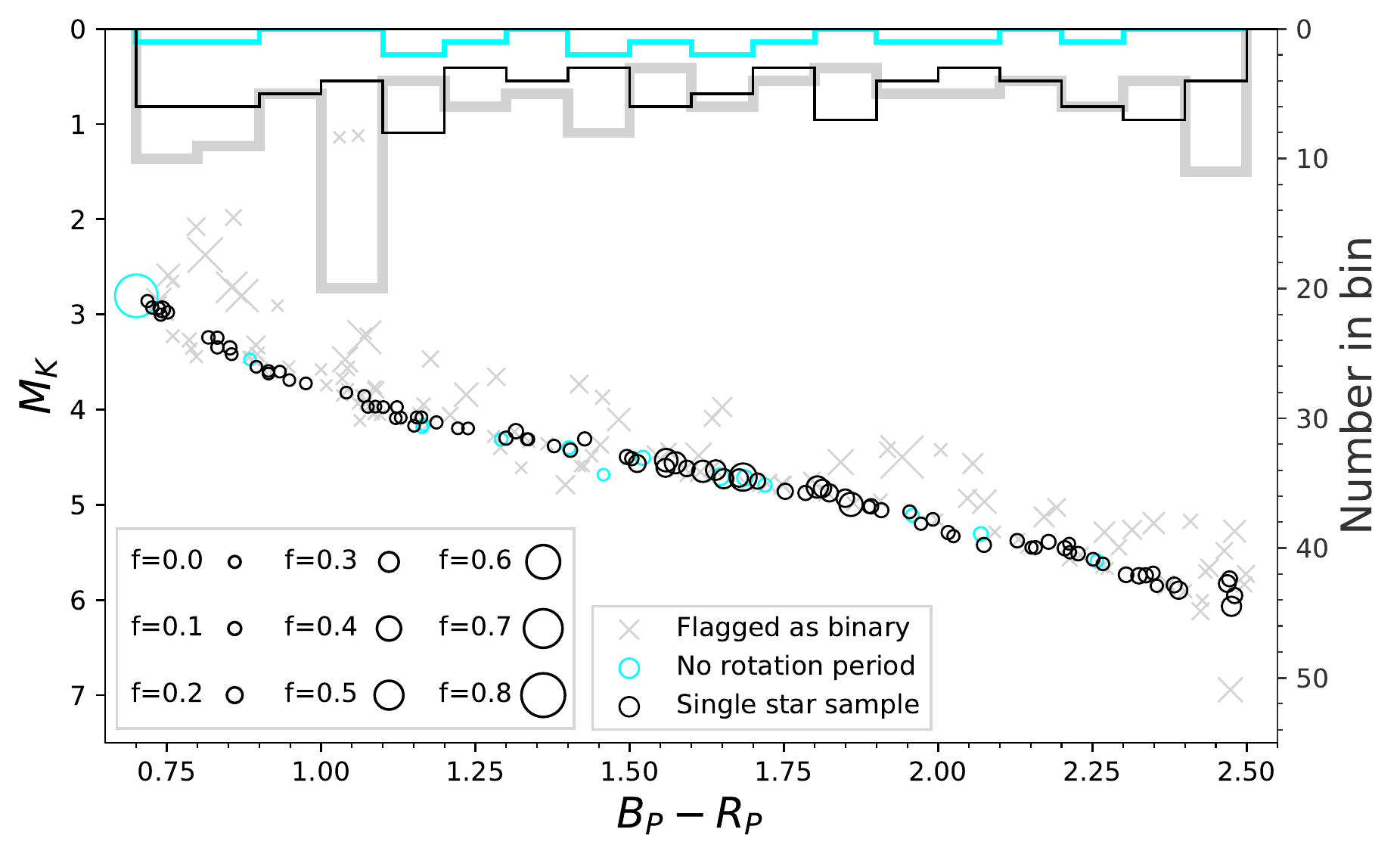}
\caption{\label{fig:RadialShears_CMD} Absolute K-band magnitude for each star plotted against Gaia $B_P-R_P$ color, points scaled according to measured starspot filling fraction. Gray crosses indicate binaries, cyan circles are stars for which there were no matches to rotation periods \citep{2017ApJ...839...92R}, and black circles are the single stars in these open cluster samples. Histograms are shown for populations of binaries (gray), stars with no reported rotation periods (blue), and single stars with measured periods (black). The rotation period completeness is high with no trends in rotation non-detections, particularly near the over-active stars ($1.5 < B_P-R_P < 1.9$).}
\end{figure}

Binarity confounds measurements of starspot filling fraction for dwarfs in the case where the companion contributes significant flux \citep{2022MNRAS.517.2165C}. In star clusters, targets with significant light from a companion appear unusually bright relative to single stars, and we identify such photometric binaries as lying more than 0.25 mag above the empirical 75th percentile main sequence. We use radial velocity and astrometric variations to detect the gravitational influence of a companion; we also identify sources which may overlap in the line-of-sight of the APOGEE fiber \citep{2022MNRAS.517.2165C}. The results of this binary rejection technique are shown in Figure \ref{fig:RadialShears_CMD}. We find 88 / 136 (65\%) of our rotating sample are single stars in Praesepe.

All single stars, including the over-active stars, pass each of the binary rejection stages where data were available; all over-active stars have Gaia astrometric {\texttt{RUWE}} measurements below 1.1, only two of them do not have an APOGEE radial velocity variability measurement, and all are on the single star photometric sequence.

The starspot measurements have been demonstrated to be sensitive even in inactive stars \citep{2022MNRAS.517.2165C}, meaning that the underlying spot measurements are unlikely to be a source of selection effects. Figure \ref{fig:RadialShears_CMD} shows a Praesepe color-magnitude diagram with a histogram indicating the distributions of binaries, stars with no detected rotation periods, and single stars in our rotation sample. There do not appear to be any significant biases in the single stars. In particular, the over-active region of stars in Praesepe is not marked by an increase in rotation non-detections, nor are there any apparently unspotted members in that temperature window. We provide spectroscopic starspot parameters and flags for this sample in Table \ref{tab:clusterdata}.

\subsection{Gyrochronal Calibration} \label{sec:gyro}
We tune the free parameters in our model to simultaneously reproduce the observed rotation of the Sun, Pleiades, and Praesepe open clusters.
From Upper Sco, we characterize the range of possible initial angular momenta as a function of mass by adopting the 7.5th, 50th, and 92.5th percentile periods in mass bins spanning 0.2--1.3 $M_\odot$ in steps of 0.1 $M_\odot$. This choice captures the range but avoids the extrema, preventing rare tidally locked binaries and background sources from dominating our fits. We draw underlying stellar structure profiles from SPOTS model grids \citep{2020ApJ...891...29S} at appropriate metallicities and spot filling fractions. These models are then subjected to a rotational braking calculation \citep{2013ApJ...776...67V} with a two-zone angular momentum transport model \citep{2010ApJ...716.1269D}. Both metallicity and starspot filling fraction affect the mapping from mass to radius and convection zone properties, which enter at significant powers in the braking law.

\begin{figure}
\centering
\includegraphics[trim={0cm 0cm 0cm 0cm},width=0.6\columnwidth]{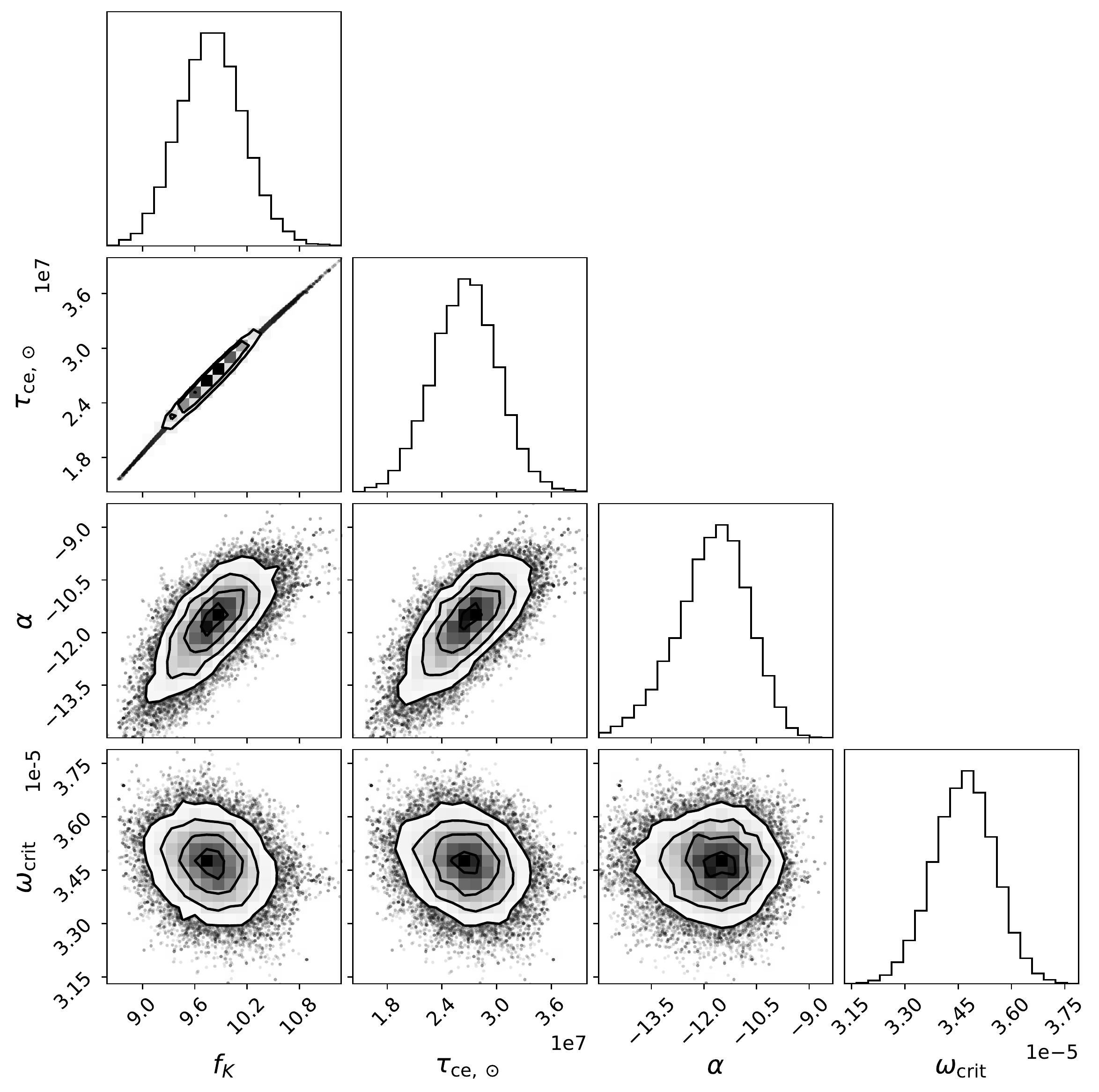}
\caption{\label{fig:RadialShears_MCMC} Marginalized 2D probability distributions for each gyrochronology fit parameter described in Appendix \ref{sec:gyro}. Convergence is seen in all parameters, with a burn-in of 300 step evaluations and a total chain length of 2100 steps. Final parameters are as follows: $f_k = 9.79 \pm 0.37$, $\omega_{\mathrm{crit}} = (3.466 \pm 0.085) \times 10^{-5}$ rad/s, $\tau_{\mathrm{ce, \,} \odot} = \left(2.66 \pm 0.36\right)\times 10^7$ yr, and $\alpha = -11.8\pm 1.0$.}
\end{figure}

We adopt the age of 112$\pm$5 Myr for the Pleiades, based on the minimum mass for the destruction of lithium in brown dwarfs \citep{2015ApJ...813..108D}. Its metallicity is near-solar, with a literature estimate of 0.03 $\pm$ 0.02 (random) $\pm$ 0.05 (systematic) dex \citep{2009AJ....138.1292S}; in this work, we assume a solar metallicity for the Pleiades.

The age of Praesepe is constrained by isochrones; for this work, we assume the value of 670$\pm$67 Myr \citep{2019ApJ...879..100D}. The metallicity of Praesepe is significantly nonsolar, with recent estimates from the literature varying from +0.11 to +0.27 \citep{2020A&A...633A..38D}. We jointly obtain a spectroscopic metallicity estimate for the cluster of +0.188$\pm$0.006 dex for 95 stars after removing two $>3\sigma$ outliers. We therefore adopt a fiducial cluster metallicity of +0.188 dex for the Praesepe gyrochrone. Changes of 0.01 dex in metallicity and 10\% in age, comparable in size to our $\sim 1\sigma$ confidence intervals, induce $\sim$5.0\% and $\sim$13\% offsets in predicted shear, which do not substantively alter our conclusions.

We divide the observed Pleiades rotation periods into 3 period bins (0--15th, 15--85th, and 85--100th percentiles) and 12 mass bins, determining the average $f_{\mathrm{spot}}$ in each bin. We initialize 12 model tracks (0.2--1.3 $M_\odot$ in steps of 0.1 $M_\odot$) using Pleiades-determined $f_{\mathrm{spot}}$ values, evolving each gyrochrone in the three Upper Sco period launch bins to Pleiades age (112 Myr adopted). For each star in each of the respective Pleiades period bins, we quantify agreement by computing $\left( P_{\mathrm{obs}} - P_{\mathrm{mod}} \right)^2 / \sigma _P ^2$ with the corresponding gyrochrone. This procedure in its entirety is repeated for [M/H] = +0.188 models launched from Upper Sco to Praesepe age (670 Myr adopted), assuming the same mass dependence of the initial rotation periods. The 50th percentile track is evaluated at solar mass, composition, age, and $f_{\mathrm{spot}}$ = 0 compared to the solar rotation period (26.870 d). From empirical relations in the Pleiades \citep{2022MNRAS.517.2165C}, we adopt a Gaussian prior on $\omega_{\mathrm{crit}}$ with a central value of $3.341\times10^{-5}$ rad/s and $1 \sigma$ range of $3.248\times10^{-5}$--$3.433\times10^{-5}$ rad/s.

The calibration of the gyrochronology models involved the evaluation of a maximum likelihood estimator at each chain step of the form $\ln P = -\frac{1}{2} \sum \left( P_{\mathrm{obs}} - P_{\mathrm{mod}} \right)^2 / \sigma _P ^2$. This was evaluated as a sum including each star in both clusters at appropriate metallicities, spot filling fractions, and across each of three gyrochrones representing each of three rotation percentile bins; model expectations were produced through linear interpolation. The Sun was also included, with an assumed error of $\sigma_P = 0.010$ d. This produces gyrochrones which produce the solar rotation rate at solar age with 50th percentile rotators to within 0.006 d. We use {\texttt{emcee}} \citep{2013PASP..125..306F} with 32 walkers run for 2100 steps each. A burn-in of 300 steps was discarded, and the subsequent chains were found to be converged using an integrated autocorrelation time diagnostic \citep{2013PASP..125..306F}.

\begin{figure}
\centering
\includegraphics[trim={0cm 0cm 0cm 0cm},width=0.8\columnwidth]{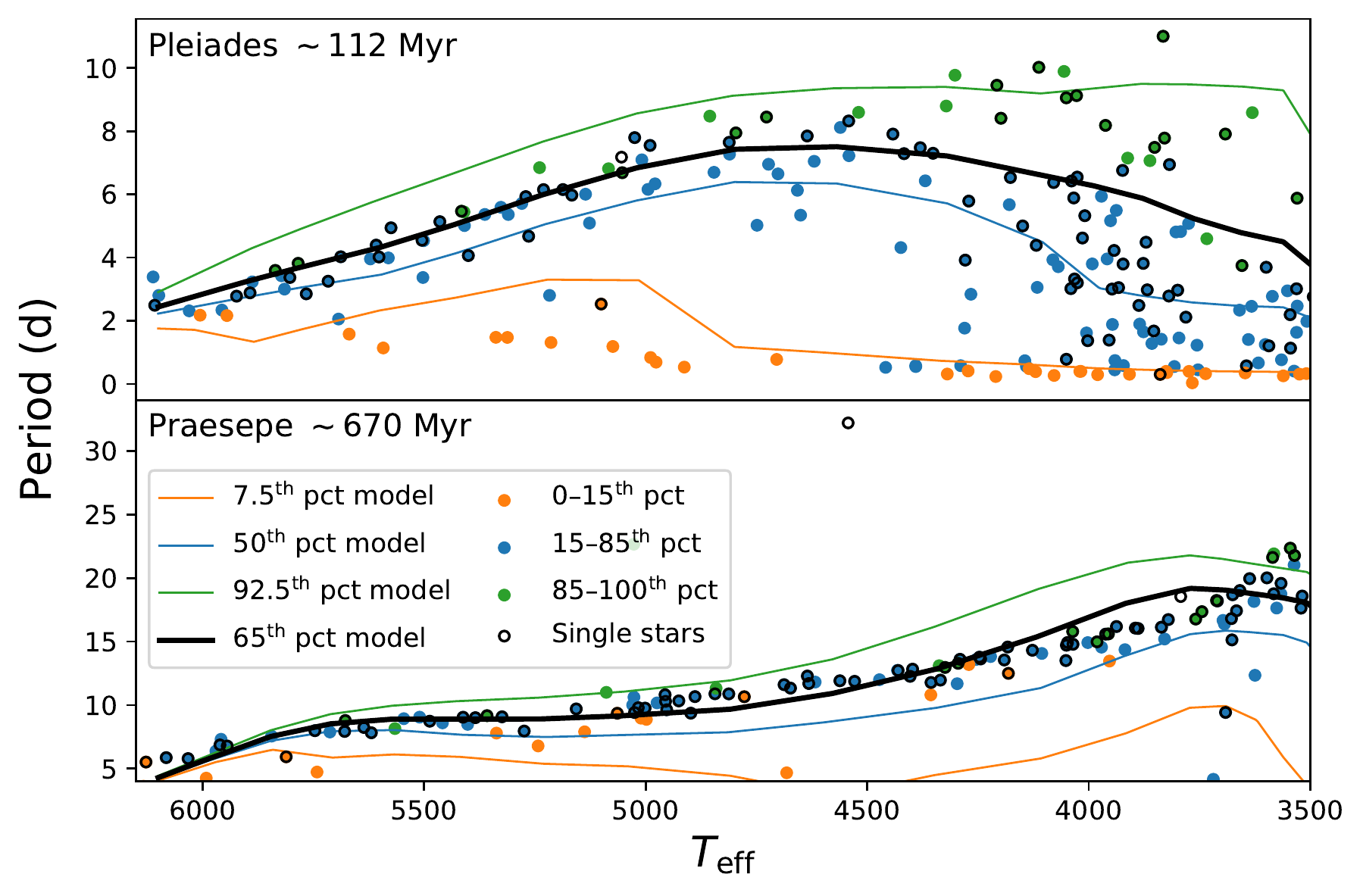}
\caption{\label{fig:RadialShears_GyroModels} Three sets of gyrochrones launched from the Upper Sco rotation distribution (at 7.5, 50, 92.5th percentiles, corresponding to orange, blue, and green, respectively) are shown in each plot against observed cluster member sequences including binaries. Observed periods are colored according to matching bins (at 0--15th, 15--85th, and 85--100th percentiles), and an additional black outline is displayed on stars which comprise our single star sample. An additional gyrochrone at the 65th percentile is shown in black against the single star sequence of each cluster. Top: the Pleiades rotation distribution and best-fit gyrochrones. Bottom: the Praesepe rotation distribution and best-fit gyrochrones.}
\end{figure}

The results from the MCMC fit are represented in Figure \ref{fig:RadialShears_MCMC}, with $\tau_{\mathrm{ce, \,} \odot} = \left(2.66 \pm 0.36 \right) \times 10^7$ yr and $\alpha = -11.8\pm1.0$. We initially find a steeper $\alpha$ than the literature, which we attribute to the inclusion of metallicities and starspots in the stellar mass determination for our fit. After accounting for these differences, our slope is consistent with previous analyses \citep{2016ApJ...829...32S}. The resultant model cluster sequences for the Pleiades and Praesepe are shown in Figure \ref{fig:RadialShears_GyroModels}. These solar-calibrated gyrochrones are also in good agreement with the observed rotation sequence of M67 (4 Gyr adopted).

\subsection{Starspots and Color Anomalies} \label{sec:coloranom}

\begin{figure}
\centering
\includegraphics[trim={0cm 0cm 0cm 0cm},width=0.6\columnwidth]{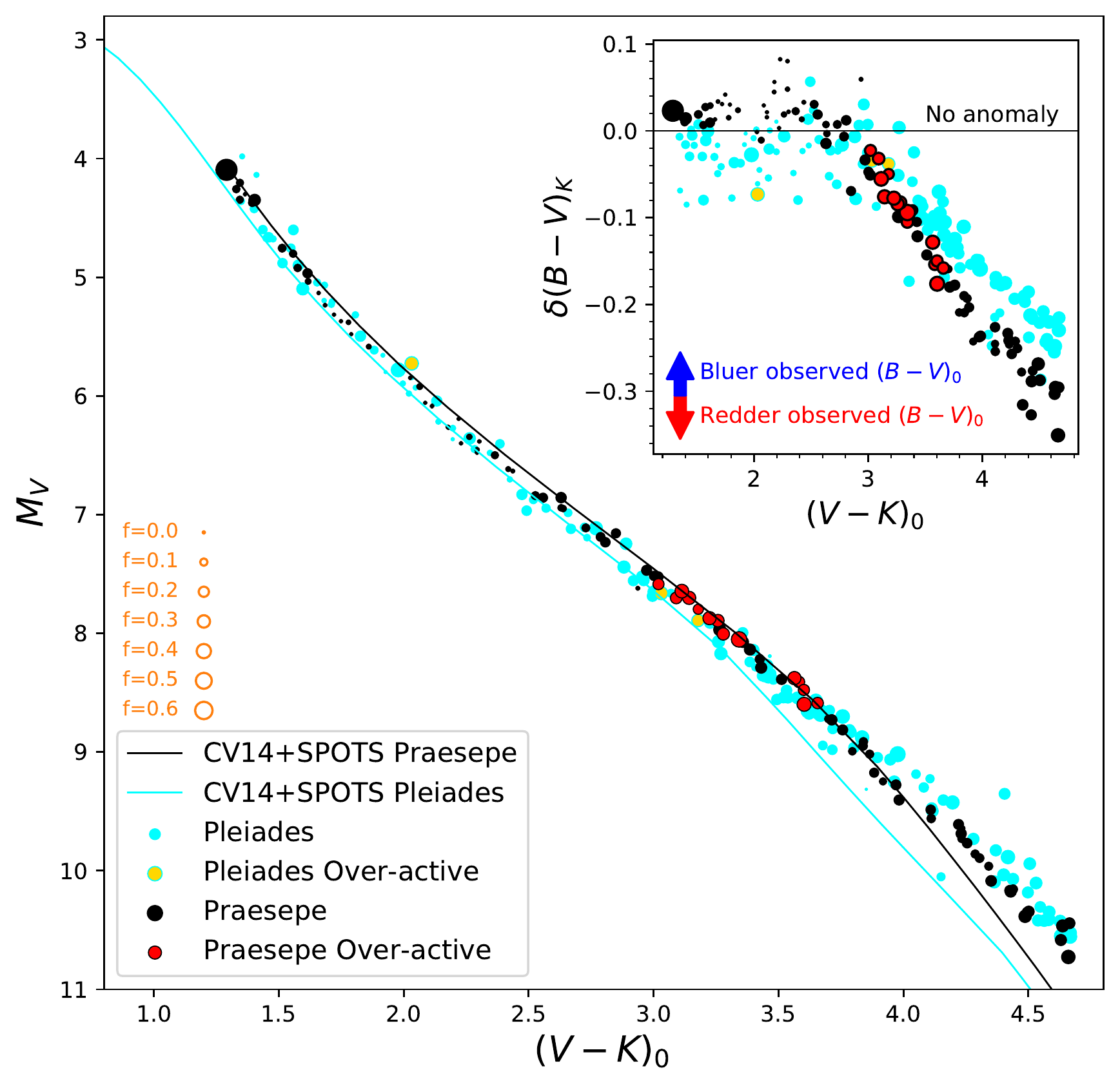}
\caption{\label{fig:RadialShears_ColorAnom} Color-magnitude diagram with Gaia DR3 synthetic photometry in dereddened $V-K$ against absolute V-band photometry. The points indicate the Pleiades (cyan), Praesepe (black), and Praesepe over-active (red) stars. The lines show standard color-magnitude relations \citep{2022MNRAS.517.2165C} for the Pleiades (cyan) and Praesepe (black). Inset: $B-V$ color anomaly evaluated at fixed K-band magnitude, plotted against dereddened $V-K$ color. Large $B-V$ color anomalies are observed to occur in stars with large surface spot filling fractions.}
\end{figure}

Color anomalies and flux excesses are known to correlate with chromospheric activity proxies and have been observed in the literature to be a result of starspots \citep{1984ApJ...283..209C, 2003AJ....126..833S}. With standard color relationships at the appropriate metallicities for the Pleiades and Praesepe \citep{2014MNRAS.444..392C}, it is possible to check whether the over-active stars appear chromospherically discrepant in photometry. In Figure \ref{fig:RadialShears_ColorAnom} we show sequences for both open clusters, showing that non-spotted stars have minimal color anomalies while all spotted stars have strongly anomalous colors. We interpret this as a photometric confirmation of the large spot fraction on these stars. The divergence between the predicted and actual color sequences at $V-K > 4$ is attributed to poor model atmosphere fits for cool stars.

\end{document}